\documentclass[fleqn,usenatbib]{mnras}

\usepackage{newtxtext}

\usepackage[T1]{fontenc}
\usepackage{ae,aecompl}
\usepackage{url}
\usepackage{multirow}
\usepackage[normalem]{ulem}
\usepackage{longtable}

\usepackage{multicol}        
\usepackage{pdflscape}  
\usepackage{deluxetable}

\usepackage{graphicx}   
\usepackage{amsmath}    
\usepackage{amssymb}    

\usepackage{color}
\newcommand{\CC}[2]{{}{#2}}

\title[FAST MSP timing]{Arecibo and FAST Timing Follow-up of twelve Millisecond Pulsars Discovered in Commensal Radio Astronomy FAST Survey}

\author[C.~C.~Miao et al.]
{C.~C.~Miao,$^{1,2}$\href{https://orcid.org/0000-0002-9441-2190}{\includegraphics[scale=0.08]{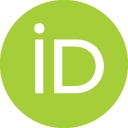}}\thanks{E-mail:miaocc@nao.ac.cn} 
W.~W.~Zhu$^{1}$\href{https://orcid.org/0000-0001-5105-4058}{\includegraphics[scale=0.08]{Figure/ORCIDiD.png}}\thanks{E-mail: zhuww@nao.cas.cn}, 
D.~Li,$^{1}$\href{https://orcid.org/0000-0003-3010-7661}{\includegraphics[scale=0.08]{Figure/ORCIDiD.png}}\thanks{E-mail: dili@nao.cas.cn}, 
P.~C.~C.~Freire$^{4}$\href{https://orcid.org/0000-0003-1307-9435}{\includegraphics[scale=0.08]{Figure/ORCIDiD.png}}, 
J.~R.~Niu$^{1,2}$\href{https://orcid.org/0000-0001-8065-4191}{\includegraphics[scale=0.08]{Figure/ORCIDiD.png}}, 
P.~Wang$^{1}$\href{https://orcid.org/0000-0002-3386-7159}{\includegraphics[scale=0.08]{Figure/ORCIDiD.png}}, 
J.~P.~Yuan$^{6}$, 
\newauthor
M.~Y.~Xue$^{1}$\href{https://orcid.org/0000-0001-8018-1830}{\includegraphics[scale=0.08]{Figure/ORCIDiD.png}}, 
A.~D.~Cameron$^{3}$\href{https://orcid.org/0000-0002-2037-4216}{\includegraphics[scale=0.08]{Figure/ORCIDiD.png}}, 
D.~J.~Champion$^{4}$\href{https://orcid.org/0000-0003-1361-7723}{\includegraphics[scale=0.08]{Figure/ORCIDiD.png}}, 
M.~Cruces$^{4}$\href{https://orcid.org/0000-0001-6804-6513}{\includegraphics[scale=0.08]{Figure/ORCIDiD.png}}, 
Y.~T.~Chen$^{1,2}$,
M.~M.~Chi$^{8}$\href{https://orcid.org/0000-0003-2650-4146}{\includegraphics[scale=0.08]{Figure/ORCIDiD.png}}, 
\newauthor
X.~F.~Cheng$^{8}$,
S.~J.~Dang$^{5}$, 
M.~F.~Ding$^{8}$,
Y.~Feng$^{7}$\href{https://orcid.org/0000-0002-0475-7479}{\includegraphics[scale=0.08]{Figure/ORCIDiD.png}}, 
Z.~Y.~Gan$^{9}$,
G.~Hobbs$^{3}$\href{https://orcid.org/0000-0003-1502-100X}{\includegraphics[scale=0.08]{Figure/ORCIDiD.png}}, 
M.~Kramer$^{4}$\href{https://orcid.org/0000-0002-4175-2271}{\includegraphics[scale=0.08]{Figure/ORCIDiD.png}}, 
Z.~J.~Liu$^{5}$,
\newauthor
Y.~X.~Li$^{9}$,
Z.~K.~Luo$^{9}$,
X.~L.~Miao$^{1}$,
L.~Q.~Meng$^{1,2}$\href{https://orcid.org/0000-0002-2885-568X}{\includegraphics[scale=0.08]{Figure/ORCIDiD.png}}, 
C.~H.~Niu$^{1}$\href{https://orcid.org/0000-0001-6651-7799}{\includegraphics[scale=0.08]{Figure/ORCIDiD.png}}, 
Z.~C.~Pan$^{1}$\href{https://orcid.org/0000-0001-7771-2864}{\includegraphics[scale=0.08]{Figure/ORCIDiD.png}}, 
L.~Qian$^{1}$\href{https://orcid.org/0000-0003-0597-0957}{\includegraphics[scale=0.08]{Figure/ORCIDiD.png}}, 
Z.~Y.~Sun$^{9}$,
\newauthor
N.~Wang$^{6}$,
S.~Q.~Wang$^{6}$,
J.~B.~Wang$^{6}$,
Q.~D.~Wu$^{6}$,
Y.~B.~Wang$^{9}$,
C.~J.~Wang$^{9}$,
H.~F.~Wang$^{10}$,
S.~Wang$^{1}$,
\newauthor
X.~Y.~Xie$^{5}$,
M.~Xie$^{8}$,
Y.~F.~Xiao$^{11}$,
M.~Yuan$^{1,2}$\href{https://orcid.org/0000-0003-1874-0800}{\includegraphics[scale=0.08]{Figure/ORCIDiD.png}}, 
Y.~L.~Yue$^{1}$\href{https://orcid.org/0000-0003-4415-2148}{\includegraphics[scale=0.08]{Figure/ORCIDiD.png}}, 
J.~M.~Yao$^{6}$,
W.~M.~Yan$^{6}$,
S.~P.~You$^{5}$,
\newauthor
X.~H.~Yu$^{5}$,
D.~Zhao$^{6}$,
R.~S.~Zhao$^{5}$,
L.~Zhang$^{1}$\href{https://orcid.org/0000-0001-8539-4237}{\includegraphics[scale=0.08]{Figure/ORCIDiD.png}} 
\\
$^{1}$National Astronomical Observatories, Chinese Academy of Sciences, 20A Datun Road, Chaoyang District, Beijing 100101, China \\ 
$^{2}$School of Astronomy and Space Science, University of Chinese Academy of Sciences, Beijing, 100049, China \\ 
$^{3}$CSIRO Astronomy and Space Science, PO Box 76, Epping, NSW 1710, Australia \\ 
$^{4}$ Max-Planck Institut f{\"u}r Radioastronomie, Auf dem H{\"u}gel 69, D-53121 Bonn, Germany\\ 
$^{5}$Guizhou Normal University, Guiyang 550001, China \\ 
$^{6}$Xinjiang Astronomical Observatory, Chinese Academy of Sciences, Urumqi, Xinjiang 830011, People's Republic of China \\ 
$^{7}$Zhejiang Lab, Hangzhou, Zhejiang 311121, People's Republic of China \\ 
$^{8}$Fudan University, Shanghai, China\\ 
$^{9}$Tencent Youtu Lab\\ 
$^{10}$College of Computer and Information, Dezhou University, Dezhou 253023\\ 
$^{11}$GuiZhou University, Guizhou 550025, China\\ 
}

\date{Accepted XXX. Received YYY; in original form ZZZ}

\pubyear{2022}

\begin{document}
\label{firstpage}
\pagerange{\pageref{firstpage}--\pageref{lastpage}}
\maketitle

\begin{abstract}
We report the phase-connected timing ephemeris, polarization pulse profiles, Faraday rotation measurements, and Rotating-Vector-Model (RVM) fitting results of twelve millisecond pulsars (MSPs) discovered with the Five-hundred-meter Aperture Spherical radio Telescope (FAST) in the Commensal radio Astronomy FAST survey (CRAFTS). The timing campaigns were carried out with FAST and Arecibo over three years. Eleven of the twelve pulsars are in neutron star - white dwarf binary systems, with orbital periods between 2.4 and 100 d. Ten of them have spin periods, companion masses, and orbital eccentricities that are consistent with the theoretical expectations for MSP - Helium white dwarf (He WD) systems. The last binary pulsar (PSR J1912$-$0952) has a significantly smaller spin frequency and a smaller companion mass, the latter could be caused by a low orbital inclination for the system. Its orbital period of 29 days is well within the range of orbital periods where some MSP - He WD systems have shown anomalous eccentricities, however, the eccentricity of PSR J1912$-$0952 is typical of what one finds for the remaining MSP - He WD systems.
\end{abstract}


\section{Introduction}
  \label{sec:intro}
Millisecond pulsars (MSP) are a special kind of neutron stars that rotate tens to hundreds of times per second and have significantly lower magnetic field strengths when compare to the normal pulsar population. 
Not only is their formation and evolution a topic of great interest, but as well their role as tools for experiments in fundamental physics -- which is achieved through pulsar timing. Examples are: 
performing tests of general relativity with binary neutron star systems \citep{tw89,Manchester:2015mda,kramer:2021ksm}, which include tests of the strong equivalence principle (SEP, \citealt{agh+18,vcf+20}) and specifically of the Lorentz invariance and conservation law \citep{Shao:2013wga,zdw+19,2020ApJ_Miao} and to measure neutron star masses, which constrain the equation of state of matter at super-nuclear density \citep{demorest+10, afw+13, Fonseca+21}, a fundamental unsolved problem in astrophysics and nuclear physics \citep{of16}. Furthermore, pulsar timing arrays, or PTAs \citep{haa+10} might soon detect low-frequency gravitational waves \citep{kbc+04}.

Since their initial discovery over 54 years ago \citep{Hewish1968}, over 3100 pulsars have been discovered and listed in Australia Telescope National Facility (ATNF) Pulsar Catalogue PSRCAT\footnote{\url{https://www.atnf.csiro.au/research/pulsar/psrcat/}} (version 1.6.5) \citep{mht+05}. The surveys that found them (like the GBT350, \citealt{snt97}, PMPS, \citealt{nlc+01}, PALFA, \citealt{cfl+06}, HTRU-S, \citealt{kjv+10}, GBNCC, \citealt{slr+14}, AO327, \citealt{dsm+13}, HTRU-N, \citealt{bck+13} and LOTAAS, \citealt{scb+19}) are driven by the need to discover all sorts of new systems: more relativistic binaries for gravity experiments, MSPs with better timing precision for PTAs, neutron stars with larger masses for probing the EOS, and the discovery of unusual systems, like the pulsar in a triple system which proved to be a superb laboratory for precise tests of the SEP. With their much higher spectral and time resolution, the more recent surveys are discovering a much larger fraction of MSPs relative to previous surveys (e.g., \citealt{pkr+19,mgf+19}), including pulsars in highly relativistic systems \citep{sfc+18,cck+18}, at least compared to any previous systems containing radio pulsars.

The Five-hundred-meter Aperture Spherical radio Telescope (FAST) \citep{nan2008, nlj+11} is currently the largest single-dish radio telescope. Early FAST commissioning observations are ongoing since August 2017 with the ultra-wideband receiver (UWB, 270-1650~MHz, Tsys~65 K) in drift scan observing mode, helping to gauge the feasibility of the drift-mode pulsar search program. The low-frequency system on FAST was decommissioned in May of 2018. We then started to implement the pilot Commensal Radio Astronomy FAST Survey (\citealt[CRAFTS\footnote{\url{https://crafts.bao.ac.cn}}]{lwq+18}) scans with the FAST 19-beam L-band (1050-1450 MHz) receiver. After January 2020, FAST finished its commissioning operation and started its science operation \citep{jth+20}. 
In January 2020, the newly formed FAST Science Committee approved five major surveys, all of which used the L-band 19beam feed-horn array. The ongoing surveys, such as the CRAFTS and the FAST Galactic Plane Pulsar Snapshot Survey \citep[GPPS]{hww+21} are expected to discover many more MSPs, and thus make a significant contribution to the PTA experiments. Until December 2021, combining all pulsars discovered in the CRAFTS and the early FAST commission survey, more than 160 new pulsars\footnote{The list of confirmed pulsars \url{https://crafts.bao.ac.cn/pulsar}} (include 35 MSPs) \citep{qpl+19, zlh+19, clh+20, wzl+21, ccl+21, wangsq2021}, 4 FRBs \citep{zll+20, nll+21} and 1 radio-faint MSP from the Fermi-LAT unassociated source \citep{wlc+21, ywy+21} have been discovered in total $\sim$ 850 hours. 
The area searched represents 15$\%$ of the sky that FAST can see.

The CRAFTS search mode observations are conducted in the drift-scan mode where the telescope is held fixed at the meridian while the sky drifts past the telescope beams. This increases the position uncertainty of the newly discovered pulsars (see \citealt{ccl+21} for more details). With the rapid growth of pulsar candidates, we strive to swiftly follow up FAST discoveries through international collaborations. Our team members from Parkes, Effelsberg, and Green Bank radio Telescopes are working on the CRAFTS pulsar follow-up \citep{clh+20, ccl+21, wzl+21}. \citealt{clh+20} reported confirmation and timing results of 11 isolated pulsars found in CRAFTS with the 64-m Parkes Radio Telescope. \citealt{ccl+21} presented confirmation and timing results of 10 CRAFTS pulsars, including 9 isolated pulsars and 1 neutron star - Carbon–Oxygen white dwarf (WD) binary system, with the 100-m Effelsberg radio-telescope. \citealt{wzl+21} showed confirmation results and nulling studies of 7 CRAFTS pulsars with the Arecibo 305-m Telescope.

In this work, we report on the timing results of twelve MSPs discovered in FAST's CRAFTS survey with FAST and also with the Arecibo 305-m radio telescope (Arecibo).
In section~\ref{sec:observation}, we describe our observation settings and epochs. The data analyses are shown in section~\ref{sec:data_analysis}. Polarization pulse profiles, timing residuals, phase-connected timing solutions, and Rotating-Vector-Model (RVM) fitting results are presented in section~\ref{sec:results}. We discuss the distribution of orbital period and eccentricity in MSP-WD systems and summarized our findings in section~\ref{sec:discussion}.

\section{OBSERVATIONS}
  \label{sec:observation}
The parameters of the FAST and Arecibo observations setup are listed in the Tab.~\ref{tab:ObsSet}.
Observations taken at Arecibo were taken with the Arecibo L-wide receiver and Puerto Rico Ultimate Pulsar Processing Instrument (PUPPI) backend. The frequency range of the Arecibo L-wide receiver was 980~MHz - 1780~MHz. Due to the RFI and low response at the edges of the band, the effective band of this receiver {\CC{is}{ was}} from 1150~MHZ to 1750~MHz. Observations at FAST are taken with the central beam of the FAST 19-beam receiver \citep{jth+20}. Its frequency coverage is 1000~MHz - 1500~MHz and the effective band is from 1050~MHz to 1450~MHz. Observations at FAST were taken by two different projects and the receiver configurations used were different (as shown in Tab.~\ref{tab:ObsSet}).
Observation data are all recorded in pulsar search mode with full-Stokes. Each pulsar observation starts with a one-minute calibration noise diode for the polarization calibration, {\CC{}{while the ﬂux calibrator observations are not performed}}. The integration time for each session of Arecibo is 1200 seconds and that of FAST is 300 seconds.

\begin{table}
  \caption{The timing observation parameters include the central frequency, bandwidth, number of channels, and sampling time.}
  \label{tab:ObsSet}
  \begin{center}
    \begin{tabular}{lcccc}
      \hline
      Telescope  & $f_{\rm{center}}$  &  $\rm{Bandwidth}$  & $n_{\rm{chan}}$ & $T_{\rm{samp}}$ \\
                 & (MHz)              &  (MHz)               &                 & ${\mu s}$  \\
      \hline
      \hline
      FAST       & 1250~MHz      &  500~MHz                  & 4096            & 49.152 \\
                 & 1250~MHz      &  500~MHz                  & 1024            & 49.152 \\
      Arecibo    & 1290~MHz      &  800~MHz                  & 2048            & 64     \\
      \hline 
    \end{tabular}
  \end{center}
\end{table}
The confirmation observations of these pulsars were made by the FAST and the regular follow-up timing observations were made by both FAST and Arecibo. In the first two weeks, 
we performed observations on days 1, 2, 3, 5, 10, and 15. After these,
we performed observations
every 30$\sim$60 days.

\begin{figure}
    \begin{center}
      \includegraphics[width= \columnwidth] {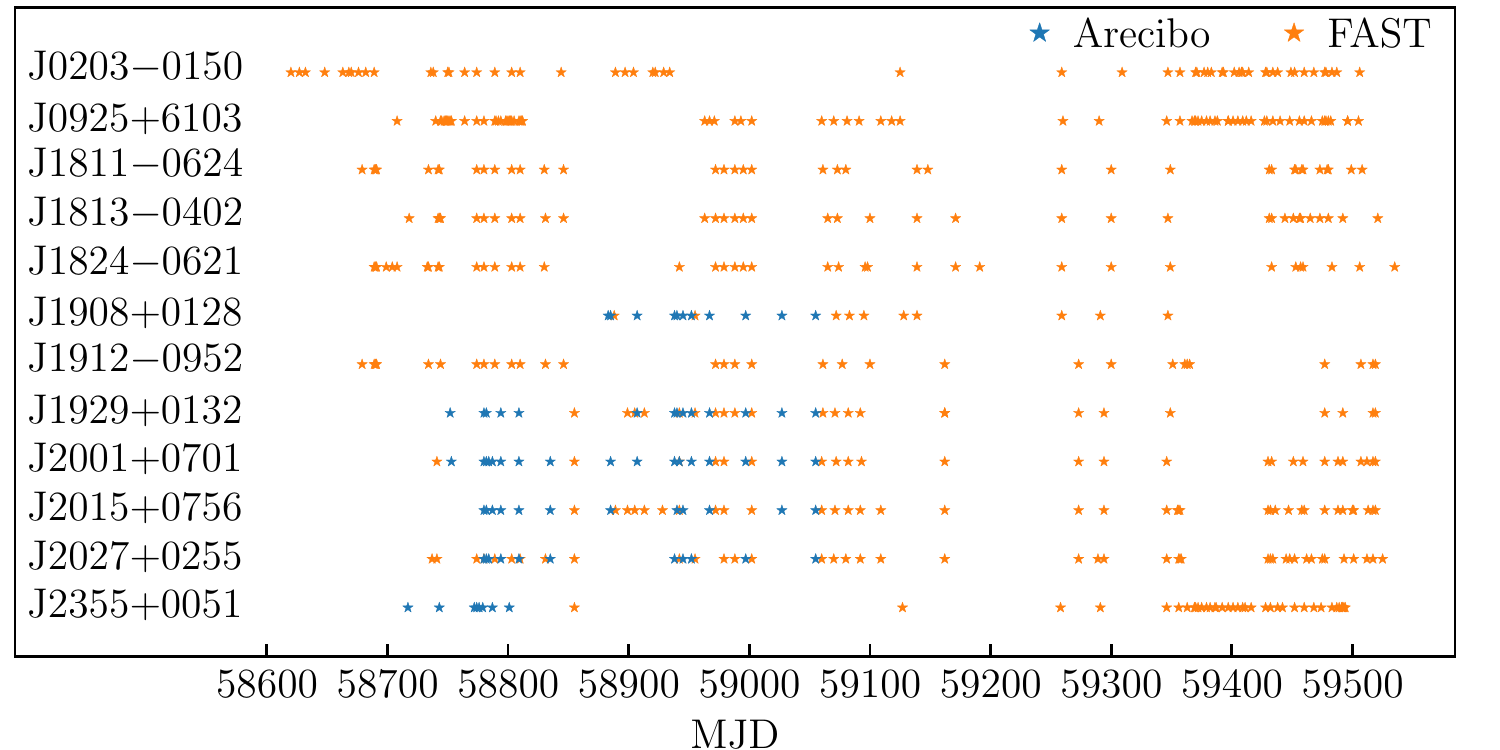}
    \end{center}
\caption{Observation epochs of 12 MSPs. The color of markers represents observation taken with different telescopes. Blue markers are Arecibo observations and orange markers are FAST observations.}
\label{fig:dataspan}
\end{figure}

Six MSPs are in Arecibo’s declination range; we followed them up with Arecibo from MJD~58883 to MJD~59055. The other six are followed up with FAST. After the collapse of Arecibo, all of the remaining timing campaigns were continued using FAST. The FAST observations were taken from MJD~58620 to MJD~59530. Epochs of observations reported in this work are shown in Fig.~\ref{fig:dataspan}. The color of the markers represents observations at the different telescopes and each marker represents each observation session.

\section{DATA ANALYSIS}
  \label{sec:data_analysis}
Unlike other new pulsar follow-up programs, where an initial ephemeris is derived for subsequent data to taken in timing and fold mode.
We take all our observation data in the incoherent searching mode. But we still need to derive an ephemeris to fold the data and derive TOAs.

To derive an initial ephemeris, we fold the raw data with an observed spin period ($\rm{P}_{\rm{obs}}$) and dispersion measure (DM) without an ephemeris. To determine these two parameters ($\rm{P}_{\rm{obs}}$, DM), we processed the first several observations with a pulsar search and analysis software, \texttt{PRESTO}\footnote{\url{https://github.com/scottransom/presto}\label{fn:presto}} (V3.0.1) \citep{ran11}. With tools from \texttt{PRESTO}, the raw data were cleaned off the radio frequency interference (RFI) and folded with an optimal $\rm{P}_{\rm{obs}}$ and DM. 
Since the initial observations are often short (5-10~min long), we could not constrain the $\dot{\rm{P}}_{\rm{obs}}$ of the pulsar from a single observation. 
From the first few months' follow-up observations results, we confirmed that only one pulsar is isolated and the other pulsars are in binary systems.

For the isolated pulsar, the initial ephemeris file contains the coordinates of the pulsar, the estimated spin frequency ($\rm{F0=1/P0}$), a fiducial spin frequency derivative ($\rm{F1}$) of 0 (to be constrained after several observations), and the measured DM at a given MJD epoch. This is often enough for fitting observations within several months and updates are needed when observations span longer than half a year. We fold the data using \texttt{DSPSR}\footnote{\url{http://dspsr.sourceforge.net}} \citep{vb11}. 
The folded data were cleaned of RFI and down-sampled in frequency with tools from \texttt{PSRCHIVE}\footnote{\url{http://psrchive.sourceforge.net}\label{fn:psrchive}} \citep{hvm04}. The times of arrival (TOAs) are created by using $\texttt{pat}^{\ref{fn:psrchive}}$.

For the pulsars in binary systems, the initial ephemeris file contains coordinates, estimated $\rm{F0}$, estimated $\rm{F1}$, DM, and the Keplerian parameters describing the orbital motion. 
Some of these parameters cannot be well constrained with few beginning observations, but have to be solved by using phase-connected TOAs from a long time span.

We use several initial observations to estimate the Keplerian binary parameters orbital period ($\rm{P_{b}}$), the epoch of periastron passage ($\rm{T_{0}}$), and projected semi-major axis ($x$) using ${\texttt{fit\_{}circular\_{}orbit.py}}^{\ref{fn:presto}}$.
The input is the optimal $\rm{F0}$ values from folding each observation with the ${\texttt{prepfold}}$ command.
${\texttt{fit\_{}circular\_{}orbit.py}}$ gives the estimation of $\rm{P_{b}}$, $\rm{T_{0}}$ and $x$ while assuming that the orbital eccentricity ($e$) is zero.

The above initial estimated circular orbit parameters are often good enough to fold the first several observations from which they were derived. 
We employ the ELL1 \citep{lcw+01, zdw+19} timing model in \texttt{TEMPO}\footnote{\url{http://tempo.sourceforge.net}} \citep{nds+15} to describe the orbit and fold the data using \texttt{DSPSR}. The folded data were processed as same as the isolated pulsar. The times of arrival (TOAs) are derived by using $\texttt{pat}$.

After we apply our initial ephemeris to fold and derive TOAs, and then iteratively improve this ephemeris with the TOAs, it quickly converges to a well-constrained ephemeris that includes only spin frequency $\rm{F0}$ and fitted circular orbital parameters. 
This new iteratively-fitted circular-orbit ephemeris is already good enough for folding most of the subsequent observations, but not good enough to phase connect them. 


To find the phase-connected timing solution to all the TOAs, we use the \texttt{TEMPO} \citep{nds+15} and \texttt{TEMPO2}\footnote{\url{https://bitbucket.org/psrsoft/tempo2}} \citep{hem06} softwares. In the final phase coherent timing solutions, we fit the pulsar's astrometry coordinates, intrinsic F0, F1, DMc and five Keplerian orbital parameters ($\rm{P_{b}}, \rm{T_{asc}}, x, \epsilon_{1}, \epsilon_{2}$) and only fit the pulsar's astrometry coordinates, intrinsic F0, F1, DM for the isolated pulsar. The integration time of each TOA is one minute for both FAST and Arecibo. We reduce the data from high signal-to-noise ratio (S/N) observations by summing their bands into 2 sub-bands and the other observations are summed into 1 sub-band. Some of our binary solutions are found with the help of \texttt{DRACULA}\footnote{\url{https://github.com/pfreire163/Dracula}} \citep{fr18}.

All of our final phase-connected timing solutions for binary systems are described by using the ELL1 model. 
This model was developed for small-eccentricity binaries and replaced $\rm{T_{0}}$, $\omega$ and $e$ by the time of ascending node ($\rm{T_{asc}}$) and the Laplace–Lagrange parameters (${\epsilon_{1} \equiv e\sin{\omega}}$, ${\epsilon_{2} \equiv e\cos{\omega}}$). The usage of $\rm{T_{asc}}$, ${\epsilon_{1}}$ and ${\epsilon_{2}}$ breaks the high correlation between $\rm{T_{0}}$ and $\omega$ in binary systems with small eccentricities and makes it a good model to connect TOAs for near-circular orbits. 
For many of these nearly-circular binaries, the ELL1 model can provide a decent description of the timing data because of the recent inclusion of $x e^2$ order terms in the description of the geometric delays in the orbit \citep{zdw+19}.

To derive the polarization profile for the pulsars, we conducted noise injection for polarimetric calibration at the beginning of each observation. After the polarimetric calibration, the highest S/N observation of each pulsar is selected and used to search the Faraday rotation measure (RM) with \texttt{RM-TOOLS}\footnote{\url{https://github.com/CIRADA-Tools/RM-Tools}} \citep{pvw+20}. The RM of ionosphere $\rm{RM_{iono}}$ are computed using \texttt{ionFR}\footnote{\url{https://github.com/csobey/ionFR}} \citep{ssh+13}. We also fitted the Rotating Vector Model \citep[RVM]{Radhakrishnan1969,Everett2001} to model the viewing geometry of these with $\texttt{psrmodel}$ from $\texttt{PSRCHIVE}^{\ref{fn:psrchive}}$. We present the best-fit RVM result when a reasonable fit can be achieved, alongside the polarimetry measurements in fig.~\ref{fig:all_profile}.

\section{RESULTS}
  \label{sec:results}

\begin{figure*}
  \begin{center}
  \includegraphics[width=2 \columnwidth] {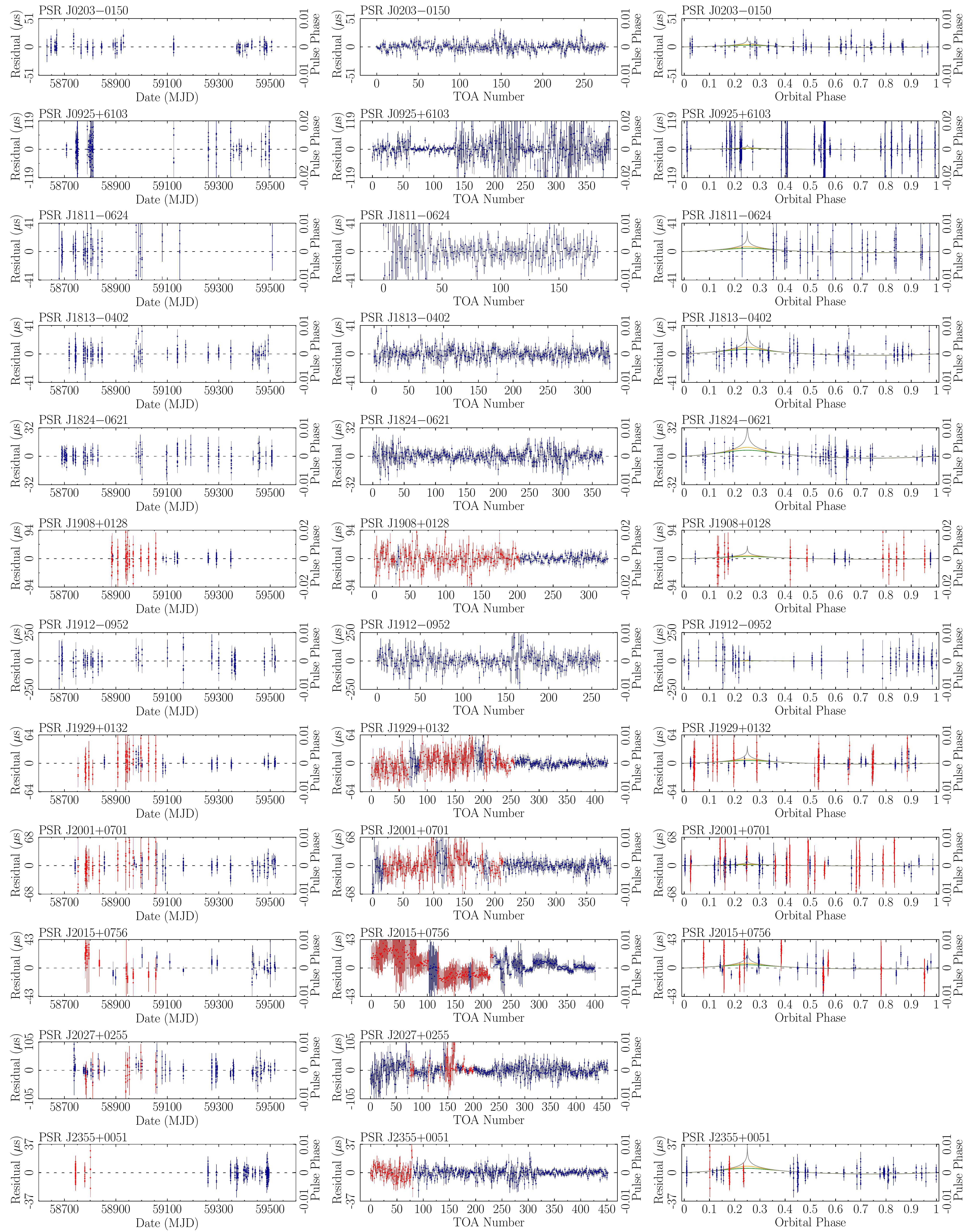}
  \end{center}
\caption{The best post-fit timing residuals of the twelve millisecond pulsars obtained using the timing solutions in Tables~\ref{tab:parTab1}-\ref{tab:parTab3}. The 3 columns are respectively MJD v.s. residual plots, TOA-number v.s. residual plots and orbital phase v.s. residual plots. The orbital phase v.s. residual plot of the only isolated pulsar J2027+0255 is not shown. The grey, orange, and green solid line shows the Shapiro delay functional form  assuming the orbital inclination angle $i=\rm{60}^{\circ}, \rm{75}^{\circ}, \rm{90}^{\circ}$. The blue and red data points represent observations with FAST and Arecibo respectively.}
\label{fig:resPlot1}
\end{figure*}

\begin{figure*}
  \begin{center}
    \includegraphics[width=2 \columnwidth]{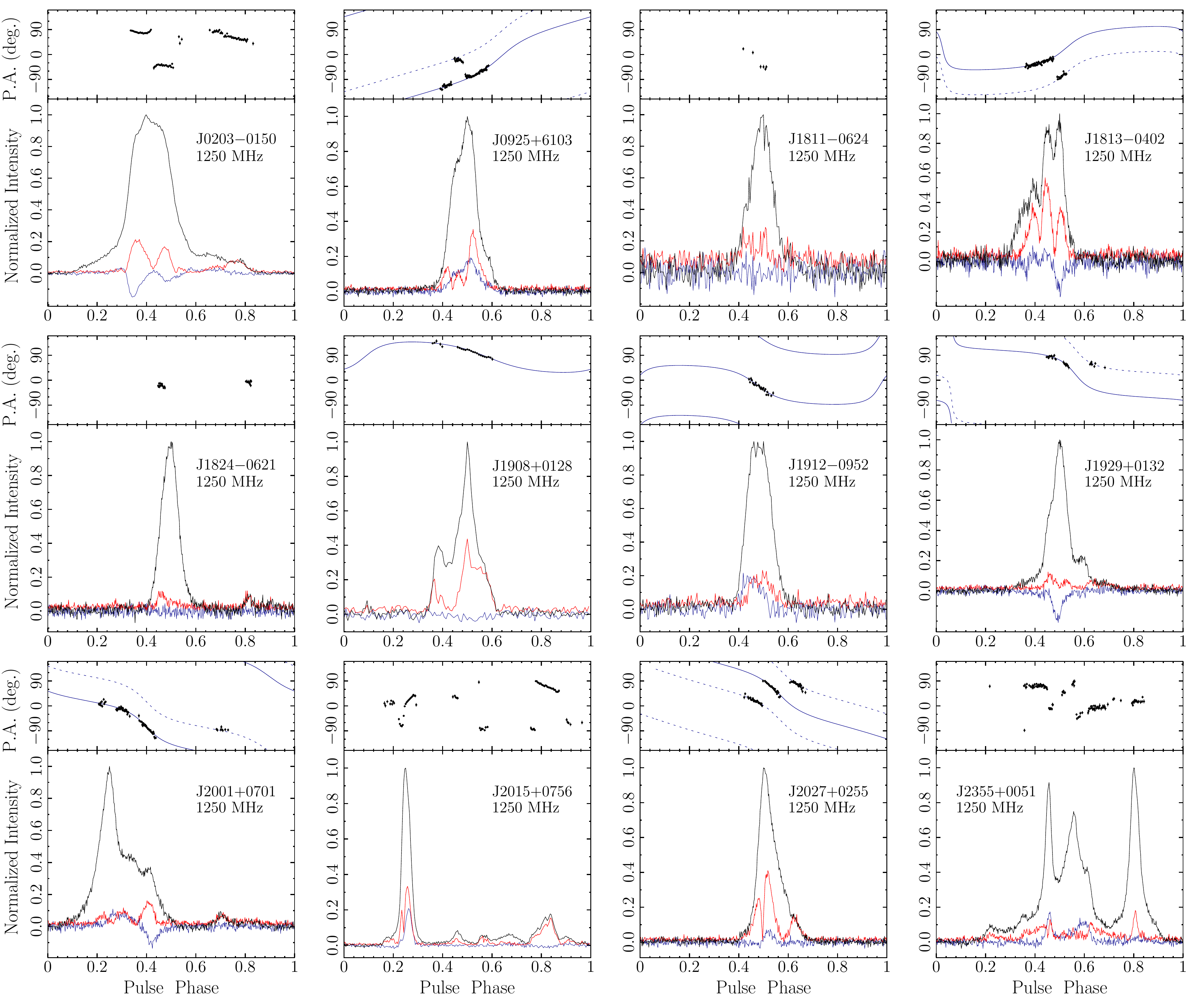}
  \end{center}
\caption{Integrated polarization profiles of 12 recycled MSPs (J0203$-$0150, J0925$+$6103, J1811$-$0624, J1813$-$0402, J1824$-$0621, J1908$+$0128, J1912$-$0952, J1929$+$0132, J2001$+$0701, J2015$+$0756, J2027$+$0255, J2355$+$0051) at 1.25 GHz. 
Top panel of each subplot: The plot of linear polarization position angle (PA) and RVM best-fit result. The black dots with error bars are the original PA points. The blue solid line represents the best-fit RVM model and the blue dashed line is $90^{\circ}$ shifted from the RVM model to indicate potential orthogonal mode. We present the RVM fitting results of 7 pulsars here. The potential orthogonal mode emission is not taken into account when fitting for the RVM model.
Bottom panel of each subplot: The plot of normalized pulsar polarization profile. The total intensity ($I$), linear polarization ($L$), and circular polarization ($V$) profiles are represented by black, red, and blue solid lines, respectively.}
\label{fig:all_profile}
\end{figure*}

\begin{table}
\caption{Faraday rotation measurement of the 12 MSPs. These measurements are made from the highest S/N follow-up observation of each pulsar. We list the measured RM ($\rm{RM_{meas}}$) and Faraday rotation from ionospheric ($\rm{RM_{iono}}$). The $\rm{RM_{iono}}$ are computed using \texttt{ionFR} \citep{ssh+13}.}
\label{tab:rm}
\begin{center}
\begin{tabular}{lcccc}
\hline
Pulsar name & Observation   & $\rm{RM_{meas}}$    & $\rm{RM_{iono}}$    & $\rm{RM_{PSR}}$      \\
            & epoch (MJD)   & (rad~$\rm{m^{-2}}$) & (rad~$\rm{m^{-2}}$) & (rad~$\rm{m^{-2}}$)  \\
\hline
\hline
J0203$-$0150  & 58844.5245  & 3.9(7)              & 1.3(1)              & 2.6(8)               \\ 
J0925$+$6103  & 58749.0424  & $-$18.1(1)          & 0.18(3)             & $-$18.28(13)         \\ 
J1811$-$0624  & 58789.3688  & 22(1)               & 1.9(1)              & 20.1(11)             \\ 
J1813$-$0402  & 58789.3826  & 24(1)               & 1.9(1)              & 22.1(11)             \\ 
J1824$-$0621  & 59506.3680  & 84.6(4)             & 3.6(2)              & 81.0(6)              \\ 
J1908$+$0128  & 59259.0495  & $-$34.7(2)          & 0.9(1)              & $-$35.6(3)           \\ 
J1912$-$0952  & 58774.4385  & $-$72.2(3)          & 3.2(2)              & $-$75.4(5)           \\ 
J1929$+$0132  & 59348.9229  & $-$69(2)            & 2.4(1)              & $-$71.4(21)          \\ 
J2001$+$0701  & 59273.1063  & $-$107.6(4)         & 1.1(1)              & $-$108.7(5)          \\ 
J2015$+$0756  & 59355.9070  & $-$70.7(6)          & 1.6(1)              & $-$72.3(7)           \\ 
J2027$+$0255  & 58774.4528  & $-$113.0(1)         & 2.7(1)              & $-$115.7(2)          \\ 
J2355$+$0051  & 59490.6882  & 4.3(6)              & 4.4(2)              & $-$0.1(8)            \\ 
\hline
\end{tabular}
\end{center}
\end{table}

\begin{table*}
\begin{center}
\caption{Best-fit \textsc{tempo} timing parameters for pulsar J0203$-$0150, J0925$+$6103, J1811$-$0624, J1813-0402.}
\label{tab:parTab1}
\begin{tabular}{lcccc}
\hline
\hline
Pulsar name  &J0203$-$0150  &J0925$+$6103  &J1811$-$0624  &J1813$-$0402   \\
\hline
\textit{Measured Parameters}&  &  &  \\
{\CC{}{Data span (MJD)}}\dotfill& {\CC{}{58632 - 59506}} & {\CC{}{58708 - 59497}} & {\CC{}{58679 - 59509}} &  {\CC{}{58718 - 59493}}\\
{\CC{}{Number of TOAs}}\dotfill&  {\CC{}{276}}&  {\CC{}{389}}&  {\CC{}{184}}&  {\CC{}{340}}\\
{\CC{}{Fit $\chi^{2}$/number of degrees of freedom}}\dotfill& {\CC{}{264.00/264}} & {\CC{}{377.18/377}} & {\CC{}{172.07/172}} &  {\CC{}{328.01/328}}\\
{\CC{}{Post-fit RMS of residuals ($\mu$s)}}\dotfill& {\CC{}{4.843}} & {\CC{}{12.530}} & {\CC{}{12.478}} &  {\CC{}{6.888}}\\
{\CC{}{EFAC}}\dotfill& {\CC{}{1.22}}& {\CC{}{1.14}}&  {\CC{}{1.12}}&  {\CC{}{1.06}}\\
Right Ascension, ${\alpha}$ (J2000) \dotfill&  02:03:33.91215(4)&  09:25:17.5451(2)&  18:11:18.0861(2)&  18:13:34.93602(2)\\
Declination, ${\delta}$ (J2000) \dotfill&  $-$01:50:01.5874(9)&  61:03:04.104(4)&  $-$06:24:37.42(3)&  $-$04:02:47.8226(12)\\
Spin Frequency, ${\nu}$ (${\text{s}^{-1}}$) \dotfill&  193.299397611600(14)&  167.15383952386(15)&  241.2147450264(12)&  243.51284588890(2)\\
Spin frequency derivative, ${\dot{\nu}}$ (${\text{s}^{-2}}$) \dotfill&  $-5.336(4)\times10^{-16}$&  $-4.51(4)\times10^{-16}$&  $-2.8(2)\times10^{-16}$&  $-4.117(7)\times10^{-16}$\\
Dispersion Measure, DM (${\text{cm}^{-3}\,\text{pc}}$) \dotfill&  19.2188(7)&  21.6442(11)&  158.826(3)&  71.3474(9)\\
{\CC{}{Flux density at 1.25 GHz (${\mu}$Jy)}}\dotfill&  {\CC{}{270(33)}}&  {\CC{}{195(59)}}&  {\CC{}{102(10)}}&  {\CC{}{148(12)}}\\
Binary model \dotfill&  ELL1&  ELL1&  ELL1&  ELL1\\
Orbital Period, ${P_{\rm b}}$ (day) \dotfill&  49.96398986(5)&  2.452162593(3)&  9.38782916(5)&  10.560352898(4)\\
Projected semi-major axis, ${x}$ (lt-s) \dotfill&  12.5803188(4)&  1.759392(3)&  6.561124(2)&  9.4263516(5)\\
Epoch of the ascending node, Tasc (MJD) \dotfill&  59395.8714674(6)&  58688.7741838(6)&  58692.1778781(8)&  58716.88174915(15)\\
${e\sin{\omega}, \epsilon_{1}( 10^{-5} )}$ \dotfill&  $-$19.375(8)&  $-$0.5(2)&  $-$0.38(5)&  $-$0.291(17)\\
${e\cos{\omega}, \epsilon_{2}( 10^{-5} )}$ \dotfill&  0.296(8)&  0.4(3)&  2.61(12)&  $-$0.297(12)\\
&  &  &  &  \\
\textit{Fixed Parameters}&  &  &  &  \\
Solar System Ephemeris \dotfill&  DE438&  DE438&  DE438&  DE438\\
Reference epoch for ${\alpha}$, ${\delta}$, and ${\nu}$ (MJD) \dotfill&  59505.767371&  59505.003027&  58743.432922&  58718.546229\\
&  &  &  &  \\
\textit{Derived Parameters}&  &  &  &  \\
Galactic longitude, ${l}$ (${^\circ}$) \dotfill&  160.317&  153.506&  22.604&  24.971\\
Galactic latitude, ${b}$ (${^\circ}$) \dotfill&  $-$59.365&  41.833&  5.923&  6.534\\
DM-derived distance (YMW16 model), ${d}$ (kpc) \dotfill&  1.84&  1.98&  5.8&  2.23\\
Spin-down luminosity, ${\dot{E}}$ (${10^{33}\,\text{erg}\,\text{s}^{-1}}$) \dotfill&  4.074(3)&  2.97(3)&  2.6(2)&  3.960(7)\\
Surface magnetic field, ${B_\text{surf}}$ (${10^{8}\,\text{G}}$) \dotfill&  2.7505(11)&  3.143(16)&  1.42(5)&  1.7087(15)\\
Characteristic age, ${\tau_\text{c}}$ (Gyr) \dotfill&  5.724(4)&  5.86(6)&  13.8(11)&  9.345(17) \\
Inferred eccentricity, e ${ ( 10^{-5} ) }$ \dotfill&  19.377(8)&  0.7(2)&  2.64(12)&  0.416(15)\\
Mass function, ${f(10^{-3}) M_{\cdot}}$ \dotfill&  0.85628165(9)&  0.972401(5)&  3.440807(4)&  8.0636184(14)\\
Minimum companion mass, $m_{\rm{c,min}} (\rm{M}_{\odot})$ \dotfill&  0.125&  0.131&  0.207&  0.283\\
Median companion mass, $m_{\rm{c,med}} (\rm{M}_{\odot})$ \dotfill&  0.146&  0.153&  0.242&  0.334\\
Longitude of the fiducial plane, $\phi_{0}$ (${^\circ}$) \dotfill&  $-$&  241(4)&  $-$&  131(1)\\
Reference position angle $\psi_{0}$ (${^\circ}$) \dotfill&  $-$&  50(5)&  $-$&  $-$56(3)\\
Magnetic inclination angle, $\alpha$ (${^\circ}$) \dotfill&  $-$&  41(15)&  $-$&  74(1)\\
Viewing angle , $\zeta$ (${^\circ}$) \dotfill&  $-$&  23(10)&  $-$&  79(2)\\
\hline
\multicolumn{5}{l}{}
\end{tabular}
\end{center}
\end{table*} 
\begin{table*}
\begin{center}
\caption{Best-fit \textsc{tempo} timing parameters for pulsar J1824$-$0621, J1908$+$0128, J1912$-$0952, J1929$+$0132.}
\label{tab:parTab2}
\begin{tabular}{lcccc}
\hline
\hline
Pulsar name  &J1824$-$0621  &J1908$+$0128  &J1912$-$0952  &J1929$+$0132   \\
\hline
\textit{Measured Parameters}\\
{\CC{}{Data span (MJD)}}\dotfill&  {\CC{}{58689 - 59507}}&  {\CC{}{58883 - 5934}}&  {\CC{}{58679 - 59520}}&  {\CC{}{58753 - 59520}}\\
{\CC{}{Number of TOAs}}\dotfill&  {\CC{}{368}}&  {\CC{}{328}}&  {\CC{}{260}}&  {\CC{}{424}}\\
{\CC{}{Fit $\chi^{2}$/number of degrees of freedom}}\dotfill&  {\CC{}{356.05/356}}&  {\CC{}{315.01/315}}&   {\CC{}{248.00/248}}&  {\CC{}{411.12/411}}\\
{\CC{}{Post-fit RMS of residuals ($\mu$s)}}\dotfill&  {\CC{}{4.960}}&  {\CC{}{13.187}}&  {\CC{}{47.834}}&  {\CC{}{8.918}}\\
{\CC{}{EFAC}}\dotfill&  {\CC{}{1.0}}&  {\CC{}{1.11}}&  {\CC{}{1.2}}&  {\CC{}{1.34}}\\
Right Ascension, ${\alpha}$ (J2000) \dotfill&  18:24:15.81954(2)&  19:08:18.80366(12)&  19:12:12.2482(3)&  19:29:42.52833(2)\\
Declination, ${\delta}$ (J2000) \dotfill&  $-$06:21:56.7918(13)&  01:28:22.714(2)&  $-$09:52:51.748(16)&  01:32:46.5630(13)\\
Spin Frequency, ${\nu}$ (${\text{s}^{-1}}$) \dotfill&  309.30868547671(2)&  212.66079094525(7)&  39.89141833615(4)&  155.80651370922(2)\\
Spin frequency derivative, ${\dot{\nu}}$ (${\text{s}^{-2}}$) \dotfill&  $-8.749(9)\times10^{-16}$&  $-2.383(5)\times10^{-15}$&  $-4.55(13)\times10^{-17}$&  $-2.078(6)\times10^{-16}$\\
Dispersion Measure, DM (${\text{cm}^{-3}\,\text{pc}}$) \dotfill&  131.0313(18)&  75.0575(17)&  126.348(15)&  53.4573(7)\\
{\CC{}{Flux density at 1.25 GHz (${\mu}$Jy)}}\dotfill&  {\CC{}{114(6)}}&  {\CC{}{143(9)}}&  {\CC{}{195(17)}}&  {\CC{}{207(20)}}\\  
Binary model \dotfill&  ELL1&  ELL1&  ELL1&  ELL1\\
Orbital Period, ${P_{\rm b}}$ (day) \dotfill&  100.91365361(6)&  84.1922026(4)&  29.4925321(5)&  9.265037044(4)\\
Projected semi-major axis, ${x}$ (lt-s) \dotfill&  44.1039113(7)&  38.482388(5)&  5.379447(7)&  8.2675144(6)\\
Epoch of the ascending node, Tasc (MJD) \dotfill&  58734.96987117(18)&  58872.1228292(12)&  58685.963815(4)&  58757.4798650(2)\\
${e\sin{\omega}, \epsilon_{1}( 10^{-5} )}$ \dotfill&  $-$22.809(2)&  7.999(17)&  $-$1.62(18)&  $-$0.812(18)\\
${e\cos{\omega}, \epsilon_{2}( 10^{-5} )}$ \dotfill&  $-$2.363(2)&  25.30(2)&  4.6(3)&  $-$0.244(18)\\
&  &  &  &  \\
\textit{Fixed Parameters}&  &  &  &  \\
Solar System Ephemeris \dotfill&  DE438&  DE438&  DE438&  DE438\\
Reference epoch for ${\alpha}$, ${\delta}$, and ${\nu}$ (MJD) \dotfill&  58780.395162&  59055.147315&  58679.659247&  58809.832979\\
&  &  &  &  \\
\textit{Derived Parameters}&  &  &  &  \\
Galactic longitude, ${l}$ (${^\circ}$) \dotfill&  24.16&  36.17&  26.415&  38.712\\
Galactic latitude, ${b}$ (${^\circ}$) \dotfill&  3.1&  $-$3.08&  $-$9.067&  $-$7.792\\
DM-derived distance (YMW16 model), ${d}$ (kpc) \dotfill&  3.53&  2.63&  9.45&  1.85\\
Spin-down luminosity, ${\dot{E}}$ (${10^{33}\,\text{erg}\,\text{s}^{-1}}$) \dotfill&  10.689(11)&  20.02(4)&  0.072(2)&  1.279(4)\\
Surface magnetic field, ${B_\text{surf}}$ (${10^{8}\,\text{G}}$) \dotfill&  1.7400(9)&  5.038(5)&  8.57(12)&  2.372(3)\\
Characteristic age, ${\tau_\text{c}}$ (Gyr) \dotfill&  5.586(6)&  1.410(3)&  13.9(4)&  11.85(3) \\
Inferred eccentricity, e ${ ( 10^{-5} ) }$ \dotfill&  22.931(2)&  26.54(2)&  4.9(3)&  0.848(18)\\
Mass function, ${f(10^{-3}) M_{\cdot}}$ \dotfill&  9.0445587(4)&  8.631737(3)&  0.1921519(7)&  7.0678361(16)\\
Minimum companion mass, $m_{\rm{c,min}} (\rm{M}_{\odot})$ \dotfill&  0.296&  0.291&  0.074&  0.270\\
Median companion mass, $m_{\rm{c,med}} (\rm{M}_{\odot})$ \dotfill&  0.349&  0.343&  0.086&  0.317\\
Longitude of the fiducial plane, $\phi_{0}$ (${^\circ}$) \dotfill&  $-$&  252(27)&  203(8)&  311(2)\\
Reference position angle $\psi_{0}$ (${^\circ}$) \dotfill&  $-$&  17(27)&  41(14)&  69(3)\\
Magnetic inclination angle, $\alpha$ (${^\circ}$) \dotfill&  $-$&  52(22)&  111(7)&  81(7)\\
Viewing angle , $\zeta$ (${^\circ}$) \dotfill&  $-$&  105(30)&  98(6)&  101(9)\\
\hline
\multicolumn{5}{l}{} \\
\end{tabular}
\end{center}
\end{table*} 
\begin{table*}
\begin{center}
\caption{Best-fit \textsc{tempo} timing parameters for pulsar J2001$+$0701, J2015$+$0756, J2027$+$0255, J2355$+$0051.}
\label{tab:parTab3}
\begin{tabular}{lcccc}
\hline
\hline
Pulsar name  &J2001$+$0701  &J2015$+$0756  &J2027$+$0255   &J2355$+$0051   \\
\hline
\textit{Measured Parameters}&  &  &  \\
{\CC{}{Data span (MJD)}}\dotfill&  {\CC{}{58741 - 59520}}&  {\CC{}{58780 - 59520}}&  {\CC{}{58737 - 59518}}&  {\CC{}{58743 - 59495}}\\
{\CC{}{Number of TOAs}}\dotfill&  {\CC{}{391}}&  {\CC{}{402}}&  {\CC{}{463}}&  {\CC{}{453}}\\
{\CC{}{Fit $\chi^{2}$/number of degrees of freedom}}\dotfill&  {\CC{}{378.00/378}}&  {\CC{}{389.08/}}&  {\CC{}{455.09/455}}&  {\CC{}{440.00/440}}\\
{\CC{}{Post-fit RMS of residuals ($\mu$s)}}\dotfill&  {\CC{}{11.444}}&  {\CC{}{6.786}}&  {\CC{}{16.097}}&  {\CC{}{3.872}}\\
{\CC{}{EFAC}}\dotfill&  {\CC{}{1.26}}&  {\CC{}{4.79}}&  {\CC{}{1.07}}&  {\CC{}{1.07}}\\
Right Ascension, ${\alpha}$ (J2000) \dotfill&  20:01:50.95896(3)&  20:15:59.59547(5)&  20:27:30.86790(7)&  23:55:51.2885(14)\\
Declination, ${\delta}$ (J2000) \dotfill&  07:01:39.6450(13)&  07:56:36.0514(16)&  02:55:37.982(3)&  00:51:09.57(4)\\
Spin Frequency, ${\nu}$ (${\text{s}^{-1}}$) \dotfill&  146.628456916751(17)&  231.05446090789(9)&  94.37951293046(4)&  268.89004281934(9)\\
Spin frequency derivative, ${\dot{\nu}}$ (${\text{s}^{-2}}$) \dotfill&  $-3.092(15)\times10^{-16}$&  $-3.56(2)\times10^{-16}$&  $-4.694(19)\times10^{-16}$&  $-2.33(3)\times10^{-16}$\\
Dispersion Measure, DM (${\text{cm}^{-3}\,\text{pc}}$) \dotfill&  63.3146(13)&  36.436(7)&  51.132(2)&  11.1452(17)\\
{\CC{}{Flux density at 1.25 GHz (${\mu}$Jy)}}\dotfill&  {\CC{}{231(18)}}&  {\CC{}{205(14)}}&  {\CC{}{186(17)}}&  {\CC{}{156(14)}}\\
Binary model \dotfill&  ELL1&  ELL1&  $-$ &  ELL1\\
Orbital Period, ${P_{\rm b}}$ (day) \dotfill&  5.22708263(2)&  6.4567191424(17)&  $-$ &  11.751022094(16)\\
Projected semi-major axis, ${x}$ (lt-s) \dotfill&  2.2302800(15)&  5.1642574(2)&  $-$ &  8.8929760(7)\\
Epoch of the ascending node, Tasc (MJD) \dotfill&  58754.6813852(11)&  58780.47407404(18)&  $-$ &  59258.1366884(2)\\
${e\sin{\omega}, \epsilon_{1}( 10^{-5} )}$ \dotfill&  $-$1.85(12)&  $-$0.130(10)&  $-$ &  $-$0.024(18)\\
${e\cos{\omega}, \epsilon_{2}( 10^{-5} )}$ \dotfill&  1.29(12)&  0.280(13)&  $-$ &  0.133(12)\\
\\
\textit{Fixed Parameters}&  &  &  \\
Solar System Ephemeris \dotfill&  DE438&  DE438&  DE438&  DE438\\
Reference epoch for ${\alpha}$, ${\delta}$, and ${\nu}$ (MJD) \dotfill&  58907.554993&  58809.870892&  58781.000000&  59492.667393\\
\\
\textit{Derived Parameters}&  &  &  \\
Galactic longitude, ${l}$ (${^\circ}$) \dotfill&  47.494&  50.126&  47.131&  95.152\\
Galactic latitude, ${b}$ (${^\circ}$) \dotfill&  $-$12.201&  $-$14.765&  $-$19.786&  $-$58.988\\
DM-derived distance (YMW16 model), ${d}$ (kpc) \dotfill&  3.78&  2.09&  4.54&  0.96\\
Spin-down luminosity, ${\dot{E}}$ (${10^{33}\,\text{erg}\,\text{s}^{-1}}$) \dotfill&  1.791(8)&  3.25(2)&  1.750(7)&  2.47(3)\\
Surface magnetic field, ${B_\text{surf}}$ (${10^{8}\,\text{G}}$) \dotfill&  3.169(7)&  1.719(5)&  7.562(15)&  1.107(8)\\
Characteristic age, ${\tau_\text{c}}$ (Gyr) \dotfill&  7.49(3)&  10.26(6)&  3.177(13)&  18.3(2) \\
Inferred eccentricity, e ${ ( 10^{-5} ) }$ \dotfill&  2.25(12)&  0.308(12)&  $-$ &  0.135(12)\\
Mass function, ${f(10^{-3}) M_{\cdot}}$ \dotfill&  0.4359286(8)&  3.5469558(5)&  $-$ &  5.4682203(14)\\
Minimum companion mass, $m_{\rm{c,min}} (\rm{M}_{\odot})$ \dotfill&  0.099&  0.209&  $-$ &  0.245\\
Median companion mass, $m_{\rm{c,med}} (\rm{M}_{\odot})$ \dotfill&  0.115&  0.245&  $-$ &  0.288\\
Longitude of the fiducial plane, $\phi_{0}$ (${^\circ}$) \dotfill&  113(1)&  $-$&  2(1)&  $-$\\
Reference position angle $\psi_{0}$ (${^\circ}$) \dotfill&  $-$80(2)&  $-$&  55(2)&  $-$\\
Magnetic inclination angle, $\alpha$ (${^\circ}$) \dotfill&  95(2)&  $-$&  124(11)&  $-$\\
Viewing angle , $\zeta$ (${^\circ}$) \dotfill&  118(2)&  $-$&  144(9)&  $-$\\
\hline
\multicolumn{5}{l}{} \\
\end{tabular}
\end{center}
\end{table*} 

Timing residuals of 2.5 years are shown in Fig.~\ref{fig:resPlot1}. To ensure the TOA residuals $\chi^{2}$ are $\sim$ 1, we weighted the TOA uncertainties by using EFAC\footnote{a scaling factor applied to TOA uncertainties}. The TOAs created from FAST observations are shown in blue and TOAs created from Arecibo observations are shown in red.

The profiles shown in Fig.~\ref{fig:all_profile} are from the observation of the highest S/N. We derive the observed RMs from these profiles and list the interstellar RMs in Tab.~\ref{tab:rm} after removing the $\rm{RM_{iono}}$. We present the RM-corrected polarization profiles at 1.25 GHz in Fig.~\ref{fig:all_profile}. We also display the linear polarization position angle (PA) of these MSPs and Rotating-Vector-Model \citep[RVM]{Radhakrishnan1969} fitting results in Fig.~\ref{fig:all_profile}. As described in \citep{Everett2001}, the linear polarization PA is a function of the magnetic inclination angle ($\alpha$), the angle between the line of sight and the rotation axis ($\zeta$), the reference position angle ($\psi_{0}$), and longitude of the fiducial plane ($\phi_{0}$). Seven MSPs in this paper with well-defined PA are successfully modeled using the RVM model and the best-fit results are listed in Tables~\ref{tab:parTab1} to \ref{tab:parTab3}. 

\begin{figure}
  \begin{center}
    \includegraphics[width= \columnwidth]{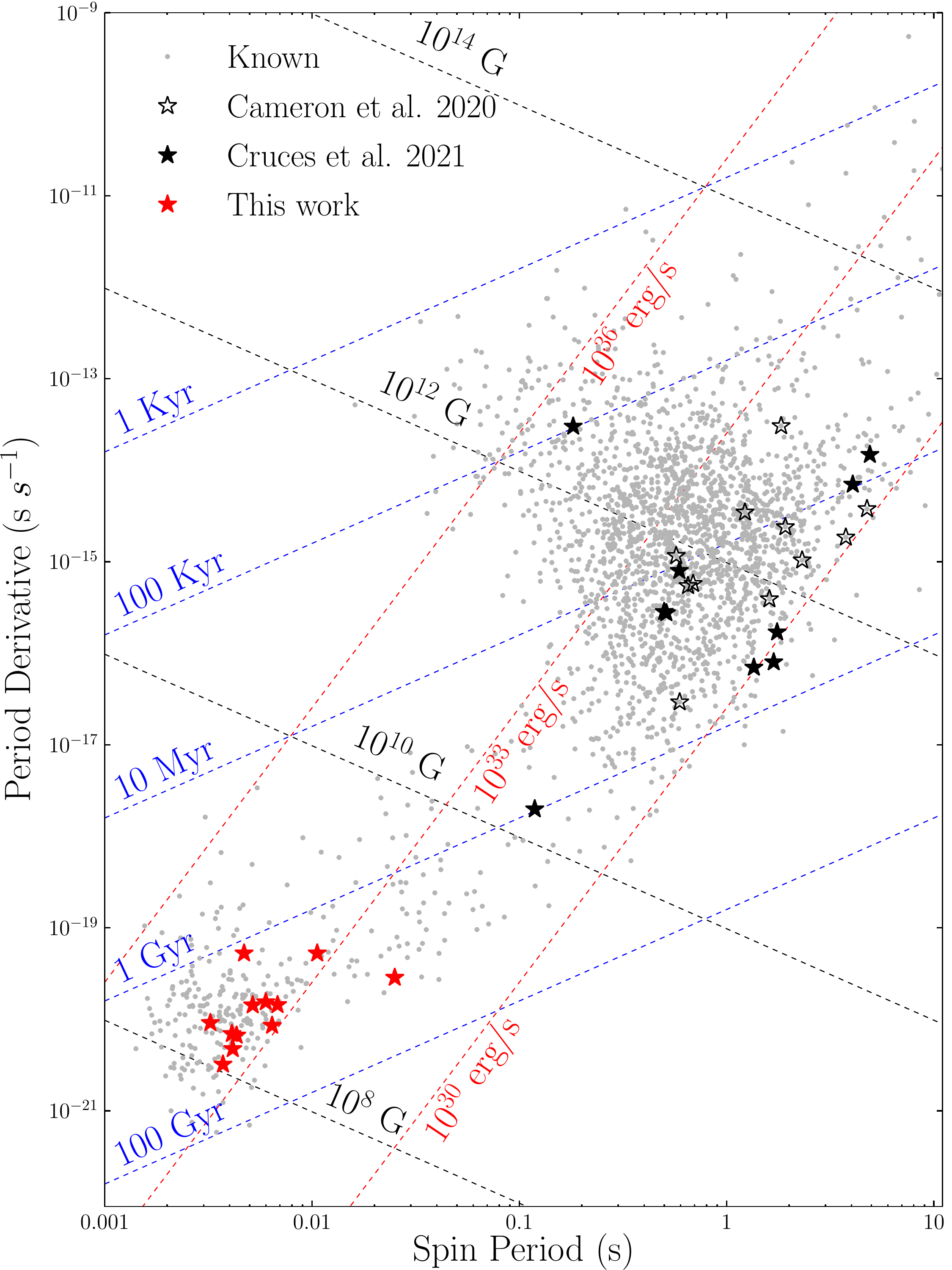}
  \end{center}
  \vspace{-0.5cm}
\caption{A $P$-$\dot{P}$ diagram of known pulsars and new pulsars presented in this work. The grey dots are 2438 pulsars listed in the \texttt{PSRCAT} (version 1.6.5) with well-measured $P$ and $\dot{P}$. The star symbols are pulsars discovered in CRAFTS with well-measured parameters. The open-black stars represent pulsars reported in \citep{clh+20} and filled-black stars represent pulsars reported in \citep{ccl+21}. The filled-red stars represent MSPs reported in this work. Lines of characteristic age (blue-dotted), surface magnetic field strength (black-dotted), and spin-down luminosity (red-dotted) are also shown in this plot.}
\label{fig:ppdot}
\end{figure}

Fig.~\ref{fig:ppdot} corresponds to a $P$-$\dot{P}$ diagram of 33 new pulsars discovered in CRAFTS with proper timing solutions and known pulsars listed in the \texttt{PSRCAT} (version 1.6.5). These newly 33 pulsars consist of 11 pulsars reported in \citep{clh+20}, 10 pulsars reported in \citep{ccl+21}, and 12 pulsars reported in this work. The $P$-$\dot{P}$ distribution of MSPs reported in this work well coincide with the distribution of known MSPs.

The timing ephemeris of these 12 MSPs are reported in Tables~\ref{tab:parTab1} to \ref{tab:parTab3}. These timing solution are obtained using JPL DE438\footnote{\url{ftp://ssd.jpl.nasa.gov/pub/eph/planets/bsp/de438.bsp}} solar system ephemeris. The latest FAST site position can be found in \citealt{qy20} and is already updated in the latest version of TEMPO (Version 13.103). 
In our analysis, we have accounted for the FAST-site to GPS clock difference.

The minimum companion mass $m_{\rm{c,min}}$ and median companion mass $m_{\rm{c,med}}$ listed in the Tabs.~\ref{tab:parTab1}, \ref{tab:parTab2} and \ref{tab:parTab3} are calculated separately assuming the orbital inclination of $i=90^{\circ}, 60^{\circ}$ and a pulsar mass of 1.4 $M_{\odot}$. We calculate the DM distances (${d}$) by using YMW 16 \citep{ymw17} models for the galactic distribution of free electrons. We estimate the spin-down luminosity (${\dot{E}}$), the surface magnetic field (${B_\text{surf}}$) and the characteristic age ${\tau_\text{c}}$ from intrinsic period (${P}$) and intrinsic spin period derivative (${\dot{P}}$) according to the equations presented in \citealt{handbook}.

Our mean flux density measurements are reported in the Tabs.~\ref{tab:parTab1}, \ref{tab:parTab2} and \ref{tab:parTab3}. For the observations taken with FAST, we estimate the pulsar flux density at 1250~MHz by using the radiometer equation \citep{handbook}.
\begin{equation}
S_{\rm{mean}} = \frac{(\rm{S/N})\beta T_{\rm{sys}}}{G\sqrt{n_{p}t_{\rm{obs}}\Delta f}}\sqrt{\frac{W}{P-W}}
\end{equation}
The gain (G) and system temperature (${T_{\rm{sys}}}$, K) of the FAST telescope are the functions of the observation zenith angle ($\theta_{\rm{ZA}}$) and following the formulas described in \citealt{jth+20}. Our observations are conducted with the 2 polarizations, an effective bandwidth (${\Delta f}$) of 400~MHz and a correction factor ${\beta}$ of 1. The equivalent width W (s), S/N are estimated from the summed pulse profile.

PSR~J0203$-$0150 is an MSP with a spin period of 5.17~ms and was first reported in \cite{yuqiuyu+20}. It is in a 49.96 day binary orbit with an eccentricity of 0.0001938. As shown in Fig.~\ref{fig:all_profile}, this MSP has an extremely wide single-component pulse profile. The $W_{10}$ pulse width (corresponding to about 10 percent of the maximum intensity) of PSR~J0203$-$0150 is 2.54~ms and the duty cycle of on pulse area is more than 70~\%. The complex PA swing of the pulsar did not yield a satisfactory RVM model fit.

PSR~J0925$+$6103 is a 5.98~ms pulsar with a DM of 21.6~${\rm {cm^{-3}~pc}}$. It is in a 2.45 day binary orbit with an eccentricity of $0.7\times10^{-5}$. We had observed this pulsar 78 times and it only showed up with enough S/N in 33 observation sessions. Within those observations with high S/N, we found no flux density variations in short time scale. The scintillation timescale for pulsars in binary systems may vary with the change of orbital phase due to modulation of the orbital velocity; however, we found that detections happen at random orbital phases, further proving that the interstellar scintillation timescale is larger than the observation length. In this case, we think the non-detections are likely caused by the strong slow-varying interstellar scintillation or the intrinsic variation of pulsar flux density.

PSR~J1811$-$0624 is a 4.14~ms pulsar at the DM of 158.8 ${\rm {cm^{-3}~pc}}$ with a low polarized, single component pulse profile. It is in a 9.38 day binary orbit with an eccentricity of $2.64\times10^{-5}$. This MSP displays a low S/N in all of our observations. The few measurable PA points make the RVM model fitting impossible.

PSR~J1813$-$0402 has a spin period of 4.107~ms, a DM of 71.3 ${\rm {cm^{-3}~pc}}$ and an estimated DM distance of 2.23 kpc. It is in a 10.56 day binary orbit with an eccentricity of $0.416\times10^{-5}$. 

PSR~J1824$-$0621 is a 3.23~ms pulsar with the DM of 131.0 ${\rm {cm^{-3}~pc}}$. It is in a 100.9 day binary orbit with an eccentricity of 0.00022931. This system has the longest orbit period from our data set. We only measured few PA points in small ranges of its spin phase and could not get an acceptable RVM model out of them.

PSR~J1908$+$0128 is a 4.702~ms pulsar with the DM of 75.05 ${\rm {cm^{-3}~pc}}$, implying a DM distance to 2.63 kpc (YWM16 model \citealt{ymw17}). As the part of a wide binary system, the orbital period of this MSP is 84.192 day and the eccentricity is $0.0002654$.

PSR~J1912$-$0952 has a spin period of 25.068~ms and a DM of 126.3 ${\rm {cm^{-3}~pc}}$. This pulsar is in an interesting binary system with a orbital period of 29.5 days and an eccentricity of $ 4.9\times10^{-5}$ (discussed in the Section \ref{sec:discussion}).

PSR~J1929$+$0132 is a 6.418~ms pulsar with the DM of 53.45~${\rm {cm^{-3}~pc}}$. It is in a 9.26 day binary orbit with an eccentricity of $0.848\times10^{-5}$. Its profile has a single component with a Gaussian-like comparative weak tail.

PSR J2001$+$0701 has a spin period of 6.82~ms and a DM of 63.3 ${\rm {cm^{-3}~pc}}$. It is in a 5.22 day binary orbit with an eccentricity of $2.25\times10^{-5}$. The profile contains a broad main pulse component and a little, divided, and nearly 100 percent linear polarized component. There is an orthogonal mode jumps between the PA swing of these two components.

PSR J2015$+$0756 is the pulsar with a spin period of 4.32~ms and a DM of 36.4 ${\rm {cm^{-3}~pc}}$. It is in a 6.45 day binary orbit with an eccentricity of $0.308\times10^{-5}$. This MSP has the most complex pulse profile in our data set. The profile is composed of one narrow main pulse and five broad components. These components are all linked by the bridge of intermediate emission. The linear and circular polarization of the main pulse is normal. The second, third, and fifth components are highly linearly polarized, while the other following components have a low level of linear and circular polarization.

PSR J2027$+$0255 is the only isolated pulsar in our paper with a spin period of 10.596~ms and a DM of 51.1 ${\rm {cm^{-3}~pc}}$. Its profile is a broad, single component pulse. The linear polarization consists of three components and there are two orthogonal mode jumps in the PA swing.

PSR~J2355$+$0051 is a 3.719~ms pulsar at the DM of 11.14 ${\rm {cm^{-3}~pc}}$. It is in a 11.75 day binary orbit with an eccentricity of $0.135\times10^{-5}$. The estimated DM distance is 0.9548 kpc the closest and the nearest pulsar in our source list. PSR~J2355$+$0051 is detected only 24 times out of 66 observation sessions. The strong interstellar scintillation or variation of pulsar flux density is likely the reason for these non-detections (the same reasons as PSR~J0925$+$6103). The unusually wide profile consists of three main pulse components and some smaller structures alongside them. We failed to get a credible RVM fit for the complex PA swing.

\section{DISCUSSION AND CONCLUSION}
  \label{sec:discussion}
\begin{figure}
  \begin{center}
    \includegraphics[width= \columnwidth]{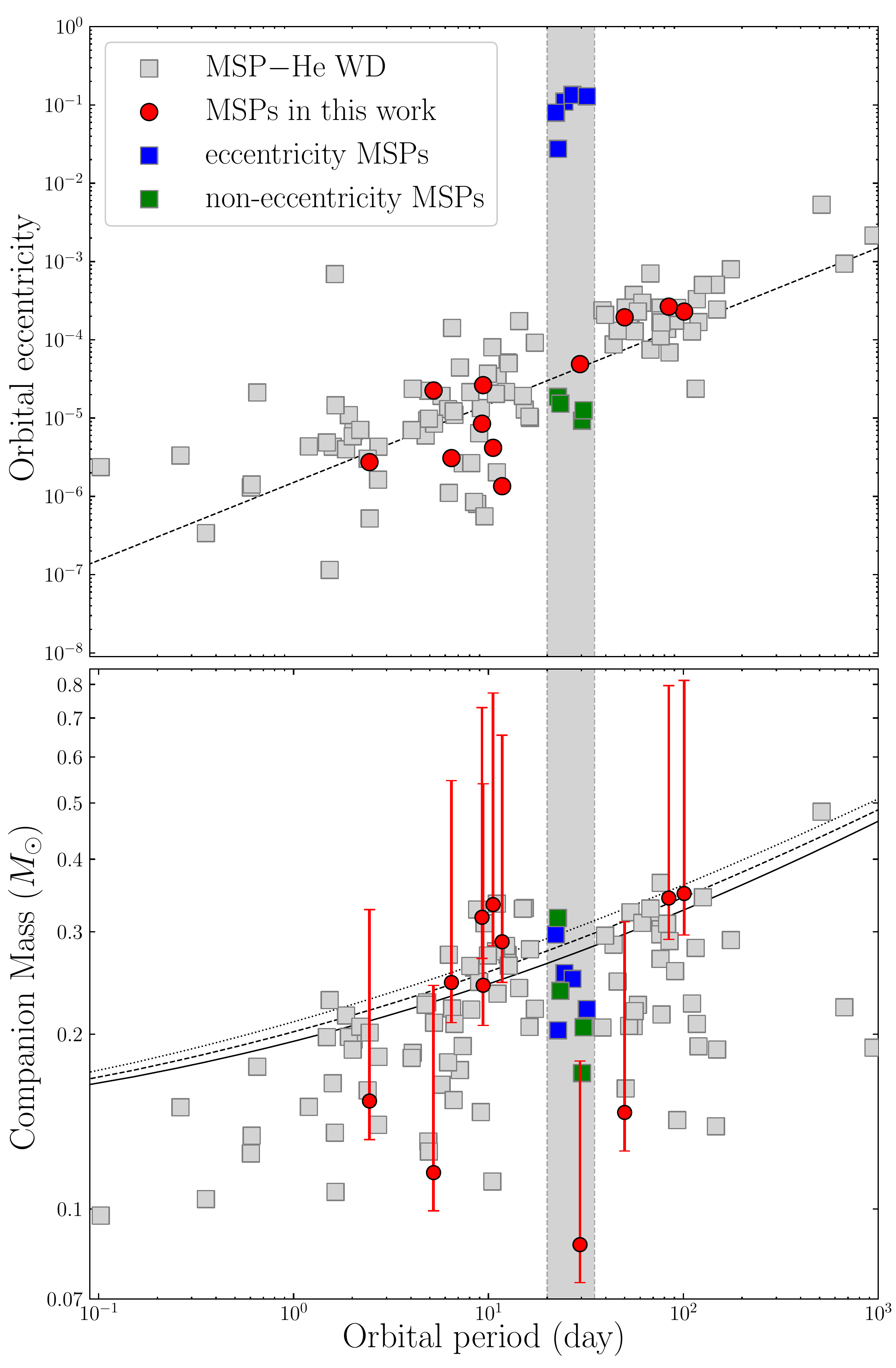}
  \end{center}
\caption{The MSP (spin periods shorter than 30 ms) with light white dwarf (WD) companions from the \texttt{PSRCAT} (version 1.6.5) \citep{mht+05}. MSP-WD systems are shown with grey squares. Eccentric systems in orbital period $P_{b}$ ranging from 20 to 35 days are marked with blue squares and non-eccentric systems in this range are marked with green squares. The red filled dots are the binaries presented in this work. The orbital period $P_{b}$ range where extra-eccentric MSP binary are often found, from 20 to 35 days is marked using shaded vertical bar.
Top panel: The $P_{b}$ and eccentricity $e$ correlation for MSP with light WD companion systems. The dashed black line shows the theoretical relation of orbital period and eccentricity given in \citealt{phin1992RSPTA}. 
Bottom panel: The $P_{b}$ and companion mass $M_{c}$ correlation for binary pulsars with light WD companion systems. The dash and solid lines are numerical calculation from \citealt{Tauris+1999} for MSP - He WD systems. Median $M_{c}$ estimates are given from their mass functions assuming an orbital inclination $i=60^{\circ}$ and $M_{psr}=1.4M_{\odot}$. The error bars are the minimum ($i = 90^{\circ}$) and 90 per cent probability ($i = 25.8^{\circ}$) $M_{c}$ estimates.}
\label{fig:pbtoecc_mc}
\end{figure}

The binaries in our sample have companion masses that are consistent with the prediction of \cite{Tauris+1999} for MSPs with Helium white dwarf (He WD) systems (see Fig.~\ref{fig:pbtoecc_mc}, bottom panel); their spin periods are also well within the typical range observed for such systems. This suggests that these systems are also MSP - He WD systems. We searched the potential counterparts at the Gaia DR2 \citep{gaiadr2+18}\footnote{\url{https://gea.esac.esa.int/archive/}} catalogs and none were found within a 1 arcsec radius. The distribution of orbital eccentricities ($e$) with $P_{b}$ (see Fig.~\ref{fig:pbtoecc_mc}, top panel) also supports this case. MSP - He WD systems in globular clusters (GCs) are excluded in our sample because their eccentricities can be greatly amplified by close encounters with other stars in those systems (for a review, see \citealt{phin1992RSPTA}).

The predictions of \citealt{phin1992RSPTA} for MSP - He WD systems are plotted in the top panel of Fig.~\ref{fig:pbtoecc_mc} with a dashed black line. The predictions of \citealt{Tauris+1999} for MSP - He WD systems are shown in the bottom panel of Fig.~\ref{fig:pbtoecc_mc} with the dashed black lines. The `orbit period gap' \citep{Tauris1996} ranging from 20 to 35 days is marked by the vertical shaded bar.

This gap is interesting for several reasons: It was originally noticed because, for a long time, no systems were known to have orbital periods within that range. Now several MSP-WD systems have been found within this gap (which has now been closed), but many are somewhat anomalous; they are known as eccentric MSP binaries (eMSPs): PSRs J0955$-$6150 \citep{crr+15, Serylak+22}, J1946$+$3417, \citep{bck+13, bfk+17}, J1950$+$2414 \citep{kls+15, zfk+19}, J1618$-$3921 \citep{ocg+18},
and J2234$+$0611 \citep{dsm+13, aks+16, sfa+19}); these system are marked in Fig.~\ref{fig:pbtoecc_mc} by the blue symbols. Although their companion masses are close to the \citealt{Tauris+1999} prediction for MSP - He WD systems, their orbital eccentricities are far from the prediction of \cite{phin1992RSPTA}. Their formation is still mysterious \citep{Tauris1996, fpt2014, hwh+18, han+21, lkf+21}.

In this gap, there are also some MSP - He WD systems that do follow the prediction of \cite{phin1992RSPTA}: PSRs J0732$+$2314 \citep{mgf+19}, J1421$-$4409 \citep{sfj+20}, J1709$+$2313 \citep{lwf+04} and J2010$+$3051 \citep{pkr+19}, these are indicated in these figures by the green symbols. Their existence shows that the process that generated the eccentricities of the eMSP systems does not do so for all MSP - He WD binaries.

Here, we present another low-eccentricity binary system in this gap. PSR~J1912$-$0952 is an MSP with an orbital period of 29.49 days and an eccentricity of $4.9\times10^{-5}$, which agrees with the relationship given by \cite{phin1992RSPTA}. This suggests that
this system likely formed through the same evolution track as the other MSP binaries \citep{phin1992RSPTA, Tauris+1999}. However, this pulsar is also slightly unusual in having a spin period of 25.07 ms and an unusually small mass function; the latter is either caused by a small inclination or
an unusually small companion mass. 

In summary, our results and discussions have shown the following:

(1) We present the timing results of 12 newly discovered pulsars from CRAFTS survey. These pulsars were followed at FAST and Arecibo at L-band. We report the phase-connected timing ephemeris, polarization pulse profiles, and Faraday rotation measurements of these new MSPs. These pulsars are all MSPs, including eleven MSP - He WD systems and one isolated pulsar.

(2) We discuss the distribution of $M_{c}$ and $e$ in our newly discovered pulsars compared with the known MSP - He WD binary systems. PSR~J1912$-$0952, PSR~J0203$-$0150, PSR~J0925$+$6103 and PSR~J2001$+$0701 are four systems with eccentricity consistent with the prediction of \citealt{phin1992RSPTA} but with $M_{c}$ lightly less than the prediction \cite{Tauris+1999}. This could be simply due to their systems having a low orbital inclination. The other seven binaries discussed in this paper have eccentricities and companion masses that are 
consistent with the prediction of \citealt{phin1992RSPTA} and \cite{Tauris+1999}.

(3) As shown in the top panel of Fig \ref{fig:pbtoecc_mc}, there exist a special out-lier group to the Phinney Orbit period and eccentricity relation -- the so-called eccentric MSPs, which fall in an `orbit period gap' \citep{Tauris1996, fpt2014, hwh+18, han+21, lkf+21}. The existence of this group indicates an alternative binary evolution channel that might have caused their unusually high orbital eccentricity.
In our sample, PSR~J1912$-$0952 is a low-eccentric MSP - He WD system ($e=4.9\times10^{-5}, P_{b} \sim 29.49 d$) that resides well in the `orbit period gap'. But its eccentricity and $P_{b}$ follows closely the prediction given in \citealt{phin1992RSPTA}. This system supports that at least some systems in this `orbit period gap' are formed through the same channel as the other MSP binaries \citep{mgf+19}.

(4) As shown in Fig. \ref{fig:resPlot1}, PSR~J1811$-$0624, PSR~J1824$-$0621, and PSR~J1908$+$0128 are three systems that have no data sets at the superior conjunction means we can’t measure Shapiro delay with the current data presented in this work.
PSR J0203-0150, PSR J2355$+$0051, PSR~J1813$-$0402, PSR~J1929$+$0131, and PSR~J2015$+$0755 are five systems with few observations at the superior conjunction. In the current data sets, the Shapiro delay gives the limitation of ${i<75^{\circ}}$.
The other four PSR~J1913$-$0952, PSR~J2001$+$0701, and PSR~J0825$+$6103 are binary systems with the timing residuals larger than the relativistic signal. The current timing precision limited the Shapiro delay measurements in these systems.

(5) Finally, we estimate that the flux density of these pulsars are greater than 0.1 mJy at 1.25 GHz. 11 of 12 MSPs have the timing rms smaller than 20 $\mu s$ (except PSR~J1912$-$0952). We expect these 11 MSPs are the potential candidates for inclusion in PTAs and could make a contribution to the PTA experiment from regular timing observations \citep{iptaDR1+16,iptaDR2+18}.

\section*{ACKNOWLEDGEMENTS}
We thank the anonymous referee and Tasha Gautam for careful review and insightful comments that significantly improved the manuscript.
This work made use of the data from FAST and Arecibo. 
FAST is a Chinese national mega-science facility, operated by National Astronomical Observatories, Chinese Academy of Sciences. 
The Arecibo Observatory was a facility of the National Science Foundation operated under cooperative agreement (\#AST-1744119) by the University of Central Florida in alliance with Universidad Ana G. M\'endez (UAGM) and Yang Enterprises (YEI), Inc.
This is work is supported by the National SKA Program of China No. 2020SKA0120200, 
the National Nature Science Foundation grant No. 
12041303, 12041304, 
11873067,
U2031117, 11903049, 
11703047, 11773041, U2031119, 12173052, 
12103013, 
12103069, 
11903049, 
the National Key R$\&$D Program of China No. 2017YFA0402600, 
the CAS-MPG LEGACY project, 
the Foundation of Science and Technology of Guizhou Province Nos. (2016)4008, (2017)5726-37,(2021)023, 
the Foundation of Guizhou Provincial Education Department (No.KY(2020)003,KY(2021)303), 
Guizhou Provincial Science and Technology Foundation (Nos. ZK[[2022]304). 
L.~Zhang is funded by ACAMAR Postdoctoral Fellowship. PW was supported by the Youth Innovation Promotion Association CAS (id. 2021055)
and the CAS Project for Young Scientists in Basic Research (grant YSBR-006).
P.~Wang and J.~M.~Yao were supported by Cultivation Project for FAST Scientific Payoff and Research Achievement of CAMS-CAS.


\section*{DATA AVAILABILITY}
Origin raw data are published by the FAST data center and can be accessed through them. For other intermediate-process data, please contact the author (miaocc@nao.cas.cn).

\bibliographystyle{mnras}
\bibliography{MSPtiming.bib} 
 
\bsp	
\label{lastpage}
\end{document}